\documentstyle[preprint,aps,prb]{revtex}
\begin{document}
\draft
\preprint{Gutzwiller wave function in 3-band Hubbard model}
\title{
Gutzwiller wave function in the three-band Hubbard model: \protect\\
A variational Monte Carlo study
}
\author{A. Oguri, T. Asahata, and S. Maekawa}
\address{
Department of Applied Physics, Nagoya University,
Nagoya 464-01, Japan
}

\date{\today}
\maketitle
\begin{abstract}
The Gutzwiller wave function for the three-band Hubbard model
on a two dimensional CuO$_2$ plane
is studied by using the variational Monte Carlo (VMC) method
in the limit $U_d \to \infty$, where $U_d$ is the Coulomb repulsion
between holes on a Cu site.
The VMC results for the energy and the fraction of $d$-holes
are compared with those of the Rice-Ueda
type Gutzwiller approximation (GA).
The difference between the VMC and the GA results
are most pronounced at half-filling,
and away from half-filling the two results agree well
in both the hole-doped and the electron-doped cases.
The doping dependence of the momentum distribution function
is also studied.
\end{abstract}

\pacs{PACS numbers: 74.20, 71.30.+h, 71.10.+x}

\narrowtext
\section{INTRODUCTION}
\label{sec:intro}
The Gutzwiller wave function (GF)
has been used for microscopic investigation of
strongly correlated electron systems such as the single-band Hubbard model,
\cite{Gutzwiller,BR,Ogawa,Vollhardt,Horsch,Yokoyama,Gros,Metzner,Gebhard,VanDongen}
the periodic Anderson model,
\cite{Ueda,Varma,Brandow,Fazekas-Brandow,Oguchi,Vulovic-Abrahams,Shiba}
and the Kondo lattice model.
\cite{Fazekas-Shiba,Fazekas-MullerHartmann,Shiba-Fazekas}
In these last years, the three-band Hubbard model
on a two dimensional CuO$_2$ plane \cite{Emery}
have been studied extensively
as a model for the high-$T_c$ superconductors.
The GF has been also applied to this model with
the Rice-Ueda \cite{Ueda} type
Gutzwiller approximation (GA), \cite{Miyake,Mayou,Sarker}
and with the variational Monte Carlo (VMC) method. \cite{Coppersmith,OA,OAS}
Although the GF is a simple wave function,
analytic evaluations of expectation values with the GF
have been succeeded only in the single-band Hubbard model
on a one dimensional chain.
\cite{Metzner,Gebhard}
In the VMC method, the expectation values
are calculated numerically
taking the square of the wave function
in the first-quantization formalism
as a probability weight.
The merit of the VMC method is that
effects of the Gutzwiller projection operator
which controls the number of doubly occupied sites
and spatial correlations inherent in a starting Fermi sea
can be taken into account exactly.
These are treated with a mean-field like approximation in the GA.
Using the VMC method, Yokoyama and Shiba\cite{Yokoyama} have shown
that the Brinkman-Rice (BR) transition \cite{BR}
does not occur in the single-band Hubbard model.
This result has been proved analytically in one dimensional case
by Metzner and Vollhardt.\cite{Metzner}

In previous papers,
we have studied the normal state properties
in the GF for the three-band Hubbard model
in the half-filled case,
\cite{OA,OAS}
where the number of holes per unit cell is $n_h=1.0$.
We have confirmed that the BR transition,
which occurs in the GA, is absent within the GF
also in the three-band Hubbard model by using the VMC method.
The purpose of this paper is to study effects of doping
on the normal state properties.
For this end, we calculate the energy
and the fraction of $d$-holes away from half-filling
both in the hole-dopped ($n_h > 1.0$)
and the electron-doped ($n_h < 1.0$) cases
using a system which contains $6 \times 6$ Cu sites.
For a comparison, we examine these quantities
also by the GA.
The results show that discrepancies of the VMC and the GA results
are most pronounced at half-filling,
and away from half-filling the two results agree well
when the doping rate $|\delta|$,
which is defined by $n_h = 1.0 + \delta$, increases.
Moreover, we study the momentum distribution function $n(k)$
using a $12 \times 12$ system.
Outside the Fermi surface $n(k)$ is nonzero
due to the strong Coulomb repulsion,
and decreases when the wave number $k$
approaches to the jump at the Fermi surface.
This behavior seems to be one of the characteristics
of the GF, and was seen also
in the GF for the single-band Hubbard model.\cite{Yokoyama,Metzner}

In Sec.\ \ref{sec:GWF},
the GF for the three-band Hubbard model is introduced,
and the outline of the VMC calculation is explained.
The VMC results for the energy and the fraction of $d$-holes
are compared with those of the GA for several doping rates
in Sec.\ \ref{sec:energy}\@.
The momentum distribution function is studied in Sec.\ \ref{sec:nk}\@.
A summary is given in Sec.\ \ref{sec:summary}\@.
In the appendix,
details of the GA calculation are given, and its relation to
the auxiliary-boson mean-field theory is discussed.

\section{GUTZWILLER WAVE FUNCTION}
\label{sec:GWF}
The Hamiltonian of the three-band Hubbard model
on a two dimensional CuO$_2$ plane is given,
{\em in a hole notation}, by
\begin{equation}
  {\cal H}  = {\cal H}_{pd}
 + \Delta_0 \left ( \widehat N_p -\widehat N_d \right )
 + \; U_d \sum_{i} n_{i\uparrow}^d n_{i\downarrow}^d \;,
   \label{eq:H}
\end{equation}
\begin{eqnarray}
  {\cal H}_{pd} &=& \; t_{pd}\ \sum_{i,\sigma}
    \Bigl [ d^{\dagger}_{i \sigma}
    \left ( \ p_{i+\hat x/2 ,\sigma} + p_{i+\hat y/2 ,\sigma}
             -p_{i-\hat x/2 ,\sigma} - p_{i-\hat y/2 ,\sigma}
    \ \right )
  +  h.c. \Bigr ]\; , \label{eq:Hpd} \\
    \widehat N_p &=& \sum_{i,\sigma}
        \left ( p^{\dagger}_{i+\hat x/2 ,\sigma}p_{i+\hat x/2 ,\sigma}
        +  p^{\dagger}_{i+\hat y/2 ,\sigma}p_{i+\hat y/2 ,\sigma} \right )
   \;, \label{eq:Np} \\
    \widehat N_d &=& \sum_{i,\sigma}
              n_{i \sigma}^d \;, \label{eq:Nd}
\end{eqnarray}
where $d^{\dagger}_{i \sigma}$  creates a hole with spin $\sigma$
in the Cu-$\, 3\, d_{x^2-y^2}$ orbital at site $i$, \,
$n_{i \sigma}^d = d^{\dagger}_{i \sigma}d_{i \sigma}$,
$p^{\dagger}_{i\pm\hat x/2 ,\sigma}$ ($p^{\dagger}_{i\pm\hat y/2 ,\sigma}$)
creates a hole in the O-$\, p_x$ (O-$\, p_y$) orbital,
and $\hat x$ ($\hat y$) is a vector in the $x$ ($y$) direction
whose length is equal to the distance
between the nearest-neighboring Cu sites.
The charge transfer energy $\Delta_0$ is written in terms of
the on-site energies for a hole
on an O site $\varepsilon_p$ and a Cu site $\varepsilon_d$ as
\, $\Delta_0 =(\varepsilon_p - \varepsilon_d)/2$\,
with \,$\varepsilon_p > \varepsilon_d$.
For convenience,
the zero of energy has been chosen at
\,$(\varepsilon_p + \varepsilon_d)/2 = 0$,
and we will use the nearest neighboring $p$-$d$ hopping matrix element $t_{pd}$
as a unit of the energy, i.e., $t_{pd} = 1$.
In this paper we concentrate on the case $U_d \to \infty$,
where $U_d$ is the Coulomb repulsion between holes on a Cu site.
In this limit doubly occupied states are prohibited on Cu sites.

The Gutzwiller wave function for the three-band Hubbard model is given by
\cite{Miyake,Mayou,Sarker,Coppersmith,OA}
     \begin{eqnarray}
      & &|\Psi (\widetilde \Delta)\rangle \ =
        \text{P}_{\text{d}}|\Phi (\widetilde \Delta)\rangle \;, \\
        \label{eq:GWF}
      & &\text{P}_{\text{d}} =
        \prod_i \left [ \ 1 - n_{i\uparrow}^d n_{i\downarrow}^d \ \right ]
        \;,
        \label{eq:Pd}
     \end{eqnarray}
where $\widetilde \Delta$ is the variational parameter,
and $|\Phi (\widetilde \Delta)\rangle$ is
a ground state of a noninteracting Hamiltonian
which contains $\widetilde \Delta$ as a charge transfer energy
\begin{equation}
  \widetilde {\cal K}= {\cal H}_{pd}
   + \widetilde \Delta \left ( \widehat N_p - \widehat N_d \right )
  \;.
\label{eq:K}
\end{equation}
The wave function $|\Psi (\widetilde \Delta)\rangle$ is an extension of
the GF for the single-band Hubbard model \cite{Gutzwiller}
and the periodic Anderson model,\cite{Ueda,Shiba}
and is constructed by projecting out the states
with doubly occupied Cu sites
from the starting Fermi sea (FS), i.e.,  $|\Phi (\widetilde \Delta)\rangle$.
The optimal value of $\widetilde \Delta$ is determined
by minimizing the energy for given
Hamiltonian parameter $\Delta_0$.
If a Hartree-Fock approximation is applied
to the Hamiltonian Eq.\ (\ref{eq:H}),
the on-site energy on a Cu site is renormalized
due to the Coulomb repulsion,
and the charge transfer energy becomes smaller than the bare value $\Delta_0$.
This Hartree-Fock picture gives a simple physical interpretation of
the variational parameter $\widetilde \Delta$.
The one-particle eigenstates of $\widetilde {\cal K}$
consist of the three bands: bonding-, antibonding-, and nonbonding-band.
When the hole concentration is \,$0 \leq n_h \leq 2$,
the bonding-band, which is the lowest band, is partially filled by holes,
and the FS is written in the form
\begin{eqnarray}
|\Phi (\widetilde \Delta)\rangle &=&
  \prod_{\scriptstyle k \in {\cal F} \atop \scriptstyle \sigma}
  \gamma^\dagger_{k \sigma}|0\rangle \;, \label{eq:Fermi} \\
  \gamma^\dagger_{k \sigma}&=& \; \sqrt{ {1\over 2}
  \left ( 1+{\widetilde \Delta \over \varepsilon_k } \right ) }
  \; d^{\dagger}_{k \sigma}  \nonumber \\
  &+& \; i\, \sqrt{ {1\over 2}
  \left ( 1-{\widetilde \Delta \over \varepsilon_k } \right ) } \
  \left ( \: {V_{xk}\over V_{k}} p^{x\dagger}_{k \sigma}
        +  {V_{yk}\over V_{k}} p^{y\dagger}_{k \sigma} \: \right )
  \; , \label{eq:gamma}
\end{eqnarray}
where \, $V_{xk} = 2\,t_{pd}\,\sin(k_x/2)$,
\, $V_{yk} = 2\,t_{pd}\,\sin(k_y/2)$,
\, $V_k = \sqrt{V_{xk}^2 + V_{yk}^2}$,
\, $\varepsilon_k = \sqrt{\widetilde \Delta^2 + V_k^2}$,
and ${\cal F}$ is a set of wave numbers in the FS.
The operator $\gamma^\dagger_{k \sigma}$
creates a hole in the bonding-band, and \, $d^{\dagger}_{k \sigma}$,
\, $p^{x\dagger}_{k \sigma}$, and \, $p^{y\dagger}_{k \sigma}$
are, respectively, the Fourier transforms
  \begin{eqnarray}
    d^{\dagger}_{k \sigma} &=& {1 \over \sqrt{N_{\text{unit}}} } \sum_i
    d^{\dagger}_{i \sigma}  \; e^{i\,q R_i} \;,
      \label{eq:ddag} \\
    p^{x\dagger}_{k \sigma} &=& {1 \over \sqrt{N_{\text{unit}}} } \sum_i
    p^{\dagger}_{i+\hat x/2 ,\sigma}  \; e^{i\,q\, (R_i+\hat x/2)} \;,
      \label{eq:pxdag} \\
    p^{y\dagger}_{k \sigma} &=& {1 \over \sqrt{N_{\text{unit}}} } \sum_i
    p^{\dagger}_{i+\hat y/2 ,\sigma}  \; e^{i\,q\, (R_i+\hat y/2)}
      \label{eq:pydag}\; ,
  \end{eqnarray}
where $N_{\text{unit}}$ is the number of the unit cells
each of which contains the three sites CuO$_2$.
The summation with respect to $i$ ($R_i$) runs over the Cu sites.

The energy expectation value, {\em per unit cell \/}, is written as
  \begin{eqnarray}
  E(\widetilde \Delta) &=& \langle {\cal H} \rangle\, /\, N_{\text{unit}}
 \nonumber \\
  &=& \, E_{pd}(\widetilde \Delta) +
  \Delta_0 \left [ n_p(\widetilde \Delta) - n_d(\widetilde \Delta) \right ] \;,
  \label{eq:eg}
  \end{eqnarray}
where
\, $E_{pd}(\widetilde \Delta ) =
\langle {\cal H}_{pd} \rangle\, /\, N_{\text{unit}}$,
\, $n_p(\widetilde \Delta ) =
\langle \widehat N_p \rangle\, /\, N_{\text{unit}}$,
\, $n_d(\widetilde \Delta ) =
\langle \widehat N_d \rangle\, /\, N_{\text{unit}}$,
and
\begin{equation}
\langle \widehat {\cal O} \rangle =
{\langle \Psi (\widetilde \Delta)|\,
\widehat {\cal O} \, |\Psi (\widetilde \Delta)\rangle
\over
\langle \Psi (\widetilde \Delta)|\Psi (\widetilde \Delta)\rangle}\;.
\end{equation}
Since the total number of holes $N_h$ is a conserved quantity,
$n_p(\widetilde \Delta)$ is written in terms of $n_d(\widetilde \Delta)$:
\, $n_p(\widetilde \Delta) = n_h - n_d(\widetilde \Delta)$
with \, $n_h = N_h\, /\, N_{\text{unit}}$.
Thus, the total energy for various $\Delta_0$
can be obtained from the data of
$E_{pd}(\widetilde \Delta )$ and $n_d(\widetilde \Delta )$.
We calculate the expectation values, $\langle \widehat {\cal O} \rangle$,
numerically by using the Monte Carlo (MC) method
taking the square of the Slater determinant
as a probability weighting.
\cite{Horsch,Yokoyama,Gros,Shiba,Ceperley}
In this method, the constraint of no doubly occupied Cu sites
and spatial correlations inherent in
$|\Phi (\widetilde \Delta)\rangle$ are treated exactly.
Typically, 20 000--70 000 samples are used for the averages,
and each sample is taken every 4--7 MC steps after 10 000 MC
steps have been discarded for the relaxation,
where one MC step means a set of $N_h$--$3N_h$ trials for new configurations.
When the acceptance ratio becomes small,
that is the case for large $\widetilde \Delta$,
we increase the number of the samples and the sampling intervals
for keeping the statistical independence of the samples.
To estimate the statistical fluctuation in the MC calculation,
we divide the total samples into 10 groups and, if needed,
show the maximum and the minimum among them as an error bar.
We use a system with $6 \times 6$ unit cells (108 sites) for
calculating $E_{pd}(\widetilde \Delta )$ and $n_d(\widetilde \Delta )$,
and use a $12 \times 12$ system for the momentum distribution function.
We choose the periodic boundary condition
along the $x$-direction, and the antiperiodic boundary condition along
the $y$-direction.
With this boundary condition, \, $|\Phi (\widetilde \Delta)\rangle$
can be determined uniquely in the half-filled case.
\cite{Yokoyama}

\section{VMC AND GA RESULTS}
\label{sec:energy}

In this section, we show the results for the doping dependence of
the energy and the fraction of $d$-holes.
The total number of holes used in the calculation is
\,$N_h = 32, 36, 40, 48,$ and $56$,
and the system size is $N_{\text{unit}}=36$.
Here the case of $n_h=36/36$ corresponds to half-filling,
and  the other cases, $\, n_h < 1.0$, and, $n_h > 1.0$,
correspond to the electron-doped and the hole-doped cases,
respectively. We use $|t_{pd}|$ as units of energy.

In Figs.\ \ref{fig:ndren} and \ref{fig:Epdren},
the expectation values of the $d$-hole fraction $n_d(\widetilde \Delta )$
and the $p$-$d$ hopping energy $E_{pd}(\widetilde \Delta )$
are shown as functions of the variational parameter $\widetilde \Delta$.
The solid circles ($\bullet$) denote the VMC results,
and the solid lines denote the corresponding expectation values
in the FS, i.e.,
\, $\langle \Phi (\widetilde \Delta) | \widehat N_d |
\Phi (\widetilde \Delta)\rangle\, /\, N_{\text{unit}}\;$ or
\, $\langle \Phi (\widetilde \Delta) | {\cal H}_{pd} |
\Phi (\widetilde \Delta)  \rangle \, /\, N_{\text{unit}}$.
Thus, the differences between the circles and the solid lines
are due to effects of the projection operator $\text{P}_{\text{d}}$
defined in Eq.\ (\ref{eq:Pd}).
The statistical fluctuations due to the MC calculation
are smaller than the size of the circles.
Both of the results,
$n_d(\widetilde \Delta )$ and $E_{pd}(\widetilde \Delta )$,
show that the difference of the GF from the FS
becomes large with increasing $n_h$.
Because the doubly occupied states on Cu sites are projected out,
the fraction of $d$-holes is smaller than that in the FS,
and $n_d(\widetilde \Delta ) \leq 1.0$.
For large $\widetilde \Delta$,
\,$n_d(\widetilde \Delta)$ tends to $n_h$ in the electron-doped case (a),
and tends to $1.0$
in the half-filled (b) and the hole-doped cases (c)--(e).
The projection operator $\text{P}_{\text{d}}$
causes the loss of the hopping energy,
so that $E_{pd}(\widetilde \Delta )$ is larger than
that in the FS.
For large $\widetilde \Delta$, \, $E_{pd}(\widetilde \Delta )$
tends to zero in proportion to $1/ \widetilde \Delta$.\cite{OA}
The hopping energy in the FS
becomes minimum at $\widetilde \Delta = 0.0$,
which becomes deep when the hole concentration $n_h$ increases.
Correspondingly, \,$E_{pd}(\widetilde \Delta )$ becomes minimum
at small negative $\widetilde \Delta$.
In Fig.\ \ref{fig:epdmin},
the value of $\widetilde \Delta$ which corresponds to the minimum
of $E_{pd}(\widetilde \Delta )$ is plotted as a function of $n_h$.
Since the restriction of the $p$-$d$ hopping
due to the projection $\text{P}_{\text{d}}$
becomes {\em relatively \/} less important
when the fraction of $d$-holes is decreased,
the minimum of \,$E_{pd}(\widetilde \Delta )$
shifts to the negative value of $\widetilde \Delta$.
The shift becomes large with increasing $n_h$.

The expectation value of the total energy $E(\widetilde \Delta)$ is calculated
substituting $E_{pd}(\widetilde \Delta)$, \,$n_d(\widetilde \Delta)$,
and \,$n_p(\widetilde \Delta) = n_h - n_d(\widetilde \Delta)$
into Eq.\ (\ref{eq:eg}). In Fig.\ \ref{fig:egren},
\, $E(\widetilde \Delta)$ in the  half-filled case
is shown as a function of the variational parameter $\widetilde \Delta$
for some given $\Delta_0$.
When $\Delta_0$ increases,
the minimum of $E(\widetilde \Delta)$ becomes broad,
and the value of $\widetilde \Delta$
which corresponds to the minimum becomes large.
This is due to the facts that
the hopping energy $E_{pd}(\widetilde \Delta)$
has a minimum at $\widetilde \Delta \simeq 0$,
and the potential energy,
which is the second term in Eq.\ (\ref{eq:eg})
\,$\Delta_0 \, [ n_p(\widetilde \Delta) - n_d(\widetilde \Delta) ]$,
is a decreasing function of $\widetilde \Delta$.
The contribution of the potential energy
to $E(\widetilde \Delta)$ in the half-filled case is larger
than that in the doped cases,
which can be seen from the value of
$n_p(\widetilde \Delta) - n_d(\widetilde \Delta)$
in the limit $\widetilde \Delta \to \infty$
\[
       \lim_{\widetilde \Delta \to \infty}
       \left [\, n_p(\widetilde \Delta) - n_d(\widetilde \Delta)\, \right ] =
         \left \{
              \begin{array}{ll}
                - n_h    & \quad \mbox{for} \quad n_h < 1.0  \\
                  n_h - 2& \quad \mbox{for} \quad n_h \geq 1.0
              \end{array}
             \right. \;.
\]
Thus, although the qualitative feature of Fig.\ \ref{fig:egren}
is not changed by doping,
the broadening of the energy minimum with increasing $\Delta_0$
is most pronounced in the half-filled case.
We have determined the minimum of $E(\widetilde \Delta)$
using a least-squares fit to a quadratic function
around the stationary point.
The energy can be determined with reasonable accuracy
because an error for the energy is proportional to $\epsilon^2$,
where $\epsilon$ is the error for the variational parameter
$\widetilde \Delta$ which corresponds to the energy minimum.
The error $\epsilon$ becomes
large when the minimum is broad.
We will show the error bar due to finite $\epsilon$
if it is necessary.
In Fig.\ \ref{fig:dopt},
the optimal value of the variational parameter,
$\widetilde \Delta_{\text{opt}}$, is shown (a) as a function of $\Delta_0$,
and (b) as a function of $n_h$.
For small $\Delta_0$, \,$\widetilde \Delta_{\text{opt}}$ is negative
and a decreasing function of $n_h$
because the total energy is mainly determined
by the hopping energy $E_{pd}(\widetilde \Delta)$.
For $\Delta_0 \gtrsim 2.0$,
\,$\widetilde \Delta_{\text{opt}}$ becomes maximum at $n_h = 1.0$ (b),
and the peak becomes sharp with increasing $\Delta_0$.
This is because the total energy is dominated
by the potential energy for large $\Delta_0$.
When $\Delta_0 \gg 1.0$, the optimal value
$\widetilde \Delta_{\text{opt}}$
is proportional to $\Delta_0$.\cite{OA}

In Figs.\ \ref{fig:eg} and \ref{fig:nd},
the total energy and the $d$-hole fraction
are shown as functions of $\Delta_0$,
where the circles ($\bullet$) denote the VMC results,
and the solid lines denote the GA results.
Details of the GA calculation are given in the appendix.
The difference between
the VMC and the GA energy is large in the half-filled case (b),
and becomes small away from half-filling
both in the electron-doped case (a) and the hole-doped cases (c)-(e).
Since the zero of energy has been chosen at
$(\varepsilon_p + \varepsilon_d )/2 = 0$,
the energy becomes maximum
when the fractions of $p$-holes and the $d$-holes
turn out to be the same, i.e., $n_d = n_p$ ( $= n_h/2$ ).
In Fig.\ \ref{fig:nd}, the error bars for $n_d$ are mainly caused by
finite $\epsilon$.
In the half-filled case (b), the BR
transition occurs in the GA at  $\Delta^{\text{c}}_0 \approx 3.35$,
and  $n_d$ becomes unity  for $\Delta_0 \geq \Delta^{\text{c}}_0$.
Although the BR transition does not occur within the GF,\cite{OA}
the VMC results for $n_d$ is close to unity for large $\Delta_0$,
which means that holes tend to localize at Cu sites.
Away from half-filling, the $d$-hole fractions
with the VMC and the GA agree well.
In Fig.\ \ref{fig:egnd}, the VMC results
are plotted as functions of $n_h$ for several $\Delta_0$,
where the lines are guide to the eye.
For small $\Delta_0$,
the energy decreases with increasing $n_h$.
However, for \,$\Delta_0 \gtrsim 3.0$,
the VMC energy becomes minimum at $n_h = 1.0$
and seems to show a cusp like behavior.
If there is a jump in the slope,
the system exhibits
a phase transition to an insulating state
with $n_d < 1.0$.
The $d$-hole fraction, correspondingly,
has a peak at $n_h = 1.0$ and
becomes close to unity when $\Delta_0$ is large,
while $n_d$ is an increasing function of $n_h$ for $\Delta_0 \lesssim 1.0$.
We note that beyond the GF the metal-insulator transition
may occur in the three-band Hubbard model,
which has been studied, for instance, by Dopf {\em et al \/}
with the quantum Monte Carlo method.\cite{Dopf}

\section{MOMENTUM DISTRIBUTION FUNCTION}
\label{sec:nk}
The momentum distribution function for holes in the bonding-band
is defined by
   \begin{equation}
   n(k) = {1 \over 2} \sum_{\sigma}
   \langle \gamma^\dagger_{k \sigma}\gamma_{k \sigma} \rangle \;.
   \label{eq:nk}
   \end{equation}
The operator $\gamma^\dagger_{k \sigma}$
defined in Eq.\ (\ref{eq:gamma})
is written in a linear combination of $d$ and $p$-holes,
and the coefficient for each orbit depends
on the variational parameter $\widetilde \Delta$.
In this definition the momentum distribution function
in the FS $|\Phi (\widetilde \Delta)  \rangle$ is given by
\[
    \langle \Phi (\widetilde \Delta)
    | \gamma^\dagger_{k \sigma}\gamma_{k \sigma} |
    \Phi (\widetilde \Delta)  \rangle =
         \left \{
              \begin{array}{ll}
                1.0   & \quad \mbox{if} \quad k \in {\cal F}  \\
                0.0   & \quad \mbox{if}  \quad k \not \in {\cal F}
              \end{array}
             \right. \;.
\]
The VMC calculation for $n(k)$
is performed in the real space using the expression which is obtained by
substituting  Eq.\ (\ref{eq:gamma}) and
the Fourier transforms Eqs.\ (\ref{eq:ddag})--(\ref{eq:pydag})
into Eq.\ (\ref{eq:nk}).
In Fig.\ \ref{fig:nk}, the momentum distribution function in
the half-filled case (a),
the hole-doped case with $n_h \simeq 1.0 + 0.833$ (b),
and the  electron-doped case with $n_h \simeq 1.0 - 0.833$ (c),
are plotted for several $\widetilde \Delta$
along the line shown in (d).
Because of the antiperiodic
boundary condition in the $y$-direction,
the line is shifted from the
usual symmetry directions in the Brillouin zone.
In the Figure, the value of $\widetilde \Delta$ is given.
The corresponding Hamiltonian parameter $\Delta_0$
is estimated from the data obtained in $6 \times 6$ systems,
and is shown in the figure caption.
The statistical fluctuations in $n(k)$
are smaller than the symbols in the figures.
Inside the Fermi surface, \, $n(k)$ is almost flat,
and the jump at the Fermi surface
becomes small when $\widetilde \Delta$ (or $\Delta_0$) is large.
The jump is finite when $1/\widetilde \Delta \neq 0$.
In the limit $\widetilde \Delta \to \infty$,
the jump vanishes in the half-filled and the hole-doped cases,
while it remains finite in the electron-doped case.
Nevertheless, in the hole-doped cases,
the momentum distribution function for $p$-holes
such as $\langle p^{x \dagger}_{k \sigma}
p^{x \phantom{\dagger}}_{k \sigma}\rangle$
has a jump which remains finite
in the limit $\widetilde \Delta \to \infty$.
These tendencies can be confirmed analytically
by expanding $\gamma^\dagger_{k \sigma}$
and $|\Psi (\widetilde \Delta) \rangle$ with respect to $1/\widetilde \Delta$,
as it was performed in Ref.\ \onlinecite{OA}.
Outside the Fermi surface, $n(k)$ becomes nonzero
due to the Coulomb repulsion.
The holes excited to the outside
are $d$-holes because the projection operator $\text{P}_{\text{d}}$
acts only for $d$-holes and commutes with the operators for $p$-holes.
The value of $n(k)$ outside the Fermi surface
decreases when $k$ approaches to the jump.
This behavior seems to be one of the characteristics
of the Gutzwiller wave function,
and was seen also in the GF for the single-band Hubbard model.
\cite{Yokoyama,Metzner}

\section{SUMMARY}
\label{sec:summary}
In summary, we have studied the normal-state properties
of the three-band Hubbard model based on a Gutzwiller wave function,
and have compared the VMC results with those of the GA.
The results for the energy and the fraction of $d$-holes
show that the discrepancies of the VMC and the GA
are most pronounced at half-filling.
Away from half-filling  the results with the two methods
agree well both in the hole-doped and the electron-doped cases.
The momentum distribution function for the holes
in the bonding-band, $n(k)$, is almost flat inside the Fermi surface.
Outside the Fermi surface  $\, n(k)$ is finite
and decreases when $k$ approaches to the jump at the Fermi surface
as that in the GF for the single-band Hubbard model.

\acknowledgments
This work was supported by a Grant-in-Aid for Scientific
Research on Priority Areas from the Ministry of Education, Science,
and Culture of Japan.

\appendix
\section {GUTZWILLER APPROXIMATION}
In this appendix, we demonstrate the calculation
of the Rice-Ueda type Gutzwiller approximation (GA). \cite{Ueda}
This approximation was applied to
the three-band Hubbard model by several authors.
Miyake {\em et al \/} \cite{Miyake} and Sarker\cite{Sarker}
studied the model with finite $U_d$,
and  Mayou {\em et al \/} \cite{Mayou} studied
the case in which both  $U_d$ and  $U_p$ are $\infty$.
Here $U_p$ is the Coulomb repulsion between holes on an O site.
These authors, however, did not investigate the Brinkman-Rice transition,
which has been studied with the auxiliary-boson method
by other authors.\cite{Grilli,Balseiro,Hirashima}
The relation between the GA and the auxiliary-boson method
is briefly reviewed here.

\subsection{Method}
The Gutzwiller wave function in the limit $U_d \to \infty \:$
Eq.\ (\ref{eq:GWF}) is generalized for the model with finite $U_d$
by introducing an additional variational parameter $g$ as
  \begin{equation}
  |\Psi (g,\widetilde \Delta)\rangle \ =
  \prod_i \left [ \ 1
           - (\, 1 - g\, ) n_{i\uparrow}^d n_{i\downarrow}^d \ \right ]
        |\Phi (\widetilde \Delta)\rangle \;.
  \label{eq:GWF1}
  \end{equation}
In the GA,
the expectation value for the $p$-$d$ hopping energy,
$\langle \Psi |{\cal H}_{pd}|\Psi \rangle / \langle \Psi | \Psi \rangle$,
is replaced by that for a noninteracting band with renormalized parameters
\, $\langle \Phi_{\text{eff}}|\sqrt{q}\,
{\cal H}_{pd}|\Phi_{\text{eff}}\rangle$.
Then the energy expectation value is given by \cite{Miyake,Sarker}
  \begin{eqnarray}
  E_{\text{GA}} &=&
   \langle \Phi_{\text{eff}}|{\cal K}_{\text{eff}}|\Phi_{\text{eff}}\rangle
  + U_d d_{\text{Cu}}\, N_{\text{unit}}  \nonumber \\
  & & + \ ( \Delta_0 - \Delta_{\text{eff}})\,
  \langle \Phi_{\text{eff}}|  ( \widehat N_p -\widehat N_d )
  |\Phi_{\text{eff}}\rangle \; ,
  \label{eq:EGA}
  \end{eqnarray}
where $d_{\text{Cu}}$ is the fraction of the doubly occupied Cu sites,
and $|\Phi_{\text{eff}}\rangle$ is a ground state
of an effective Hamiltonian
  \begin{equation}
  {\cal K}_{\text{eff}} = \sqrt{q} \ {\cal H}_{pd}
    + \Delta_{\text{eff}}\  ( \widehat N_p -\widehat N_d )\; ,
  \label{eq:Keff}
  \end{equation}
with
  \begin{equation}
   q = {(n_d/2 - d_{\text{Cu}}) \over (n_d/2)(1-n_d/2)}
  \left [\, \sqrt{1 - n_d + d_{\text{Cu}}}
  + \sqrt{d_{\text{Cu}}} \  \right ]^2 \;.
  \label{eq:Q}
  \end{equation}
In deriving Eq.\ (\ref{eq:EGA}),
a configurational dependence
in the starting Fermi sea $|\Phi (\widetilde \Delta )\rangle$
defined in Eq.\ (\ref{eq:Fermi}) is neglected,
and a dominant-term approximation is used
for $d_{\text{Cu}}$ and $n_d$. \cite{Ueda,Vulovic-Abrahams}
The  condition for the dominant-term
with respect to $d_{\text{Cu}}$ is given by
   \begin{equation}
    g^2 = {      d_{\text{Cu}} (1 - n_d - d_{\text{Cu}})
           \over (n_d/2 - d_{\text{Cu}})^2   } \;.
   \label{eq:g2}
   \end{equation}
Thus the original parameter $g$ is written
as a function of $n_d$ and $d_{\text{Cu}}$.
Since the condition for the dominant-term
with respect to $n_d$ can not be written
in a simple analytic form, \cite{Vulovic-Abrahams}
the other original parameter $\widetilde \Delta$
cannot be determined within the GA.
The parameter $\Delta_{\text{eff}}$
in Eqs.\ (\ref{eq:EGA}) and (\ref{eq:Keff})
is introduced in order to treat
this condition approximately,
and is different from the original parameter $\widetilde \Delta$.
The value of  $\Delta_{\text{eff}}$ and $d_{\text{Cu}}$
are determined by minimizing the energy functional $E_{\text{GA}}$
for given $\Delta_0$ and $U_d$.

The effective Hamiltonian ${\cal K}_{\text{eff}}$ can be easily diagonalized,
and the one-particle eigenstates depend only on the ratio of
$\Delta_{\text{eff}}$ to the renormalized value of
the transfer integral $\sqrt{q}\, |t_{pd}|$
\begin{equation}
\label{eq:alpha}
\alpha = \Delta_{\text{eff}} /  \sqrt{q}\, |t_{pd}|\;.
\end{equation}
For $0 \leq n_h \leq 2$, the $d$-hole fraction,
\,$n_d = \langle \Phi_{\text{eff}}|\widehat N_d|\Phi_{\text{eff}} \rangle
/ N_{\text{unit}}$,
and the ground state energy of the effective Hamiltonian,
\, $\kappa_{\text{eff}} =
\langle \Phi_{\text{eff}}|{\cal K}_{\text{eff}}|\Phi_{\text{eff}}\rangle
/N_{\text{unit}}$,
are given by
  \begin{eqnarray}
 n_d &=& {1 \over N_{\text{unit}} } \sum_{k \in {\cal F}}
         \left ( 1 + { \Delta_{\text{eff}} \over
         \sqrt{\Delta_{\text{eff}}^2 \ + q \, V_k^2} } \right )  \nonumber \\
     & & \nonumber \\
     &=& {1 \over 2 }n_h +  \int_{-1}^{\nu_c}\! d\nu \ \rho (\nu)
         { \alpha \over \sqrt{ \alpha^2 + 4(1-\nu) }} \;, \label{eq:ndga}
  \end{eqnarray}
and
  \begin{eqnarray}
  \kappa_{\text{eff}}
  &=& - {2 \over N_{\text{unit}} } \sum_{k \in {\cal F}}
          \sqrt{\Delta_{\text{eff}}^2 \ + q \, V_k^2} \nonumber \\
  & & \nonumber \\
  &=& -2 \sqrt{q} \: |t_{pd}| \int_{-1}^{\nu_c}\! d\nu \ \rho (\nu)
                \sqrt{ \alpha^2 + 4(1-\nu)}\;, \label{eq:Egk}
  \end{eqnarray}
where  $\rho (\nu) = \sum_k \delta (\nu - \nu_k) / N_{\text{unit}}$, \,
with $\nu_k = ( \cos k_x +\cos k_y )/2$.
The parameter $\nu_c$ is determined by the hole concentration $n_h$ as
   \begin{equation}
     n_h = 2 \int_{-1}^{\nu_c}\! d\nu \ \rho (\nu) \;,
      \label{eq:chem}
   \end{equation}
and $\rho (\nu)$ is written
using the complete elliptic integral
of the first kind $K(x)$ as
$\rho (\nu) =  2\: K(\sqrt{1 - \nu^2} \,) /\pi^2$.
Substituting Eq.\ (\ref{eq:ndga}) into Eq.\ (\ref{eq:Q}),
the factor \, $q$ is written
as a function of $\alpha$ and $d_{\text{Cu}}$.
Then other quantities $\Delta_{\text{eff}}$, $\kappa_{\text{eff}}$
and $E_{\text{GA}}$ are also written
as functions of $\alpha$ and $d_{\text{Cu}}$
using Eqs.\ (\ref{eq:alpha}),
(\ref{eq:Egk}), and  Eq.\ (\ref{eq:EGA})
with $q = q(\alpha, d_{\text{Cu}})$.
When  \, $0 \leq n_h \leq 2$,
the conditions for the energy minimum,
$\partial E_{\text{GA}}/\partial \alpha =0$ and
$\partial E_{\text{GA}}/\partial d_{\text{Cu}} =0$,
are obtained as
  \begin{eqnarray}
  \Delta_0 &=& \Delta_{\text{eff}} -
     {1 \over 2}{\partial q \over \partial n_d}\;  T
 \;,\label{eq:dif-nd} \\
  U_d &=&
     {\partial q \over \partial d_{\text{Cu}}}\;  T
\;,\label{eq:dif-db}
 \end{eqnarray}
with
\begin{eqnarray}
   T  = \int_{-1}^{\nu_c}\! d\nu \ \rho (\nu)
   {4(1-\nu)t_{pd}^2 \over \sqrt{ \Delta_{\text{eff}}^2 + 4q(1-\nu)t_{pd}^2 }}
   \;.\label{eq:W}
 \end{eqnarray}

\subsection{Results for $U_d \to \infty$}

In the limit $U_d \to \infty$, the doubly occupied states are prohibited
on Cu sites,
and the factor $q$ is written substituting
$d_{\text{Cu}} \equiv 0$ in to Eq.\ (\ref{eq:Q}) as
  \begin{equation}
   q = {1 - n_d \over 1 - n_d / 2 } \;.
  \label{eq:Qinf}
  \end{equation}
The optimal value of the ratio $\alpha$ is
determined by solving Eq.\ (\ref{eq:dif-nd}).
We have calculated the integrals in Eqs.\ (\ref{eq:ndga})--(\ref{eq:chem})
numerically, and obtained the GA results
for energy and the $d$-hole fraction
shown in  Figs.\ \ref{fig:eg} and \ref{fig:nd}.
In Fig.\ \ref{fig:Q}, results for $q$ and $\Delta_{\text{eff}}$
are shown as functions of $\Delta_0$ for several $n_h$.
At half-filling, the Brinkman-Rice transition occurs
at $\Delta_0 = \Delta^{\text{c}}_0$ ($\approx 3.35$ in units of $|t_{pd}|$).
The analytic expression for $\Delta^{\text{c}}_0$ can be obtained by expanding
$E_{\text{GA}}$ with respect to $1/\alpha$ as
  \begin{equation}
   \Delta^{\text{c}}_0 = 2 \sqrt{2} \sqrt{1+(2/\pi)^2}\:|t_{pd}|\:.
   \label{eq:d0c}
  \end{equation}
For $\Delta_0 \geq \Delta^{\text{c}}_0$,
\, $\Delta_{\text{eff}}$ is a two-valued function of $\Delta_0$
when $n_h=1.0$.
The value of $\Delta_{\text{eff}}$ for $n_h=1\pm 0$,
$\Delta_{\text{eff}}^{\pm 0}$,
can be obtained  by substituting $n_d = 1.0$ and $\nu_c = 0.0$
into  Eq.\ (\ref{eq:dif-nd}) as
  \begin{equation}
   \Delta_{\text{eff}}^{\pm 0} =
   \left ( \Delta^{\text{c}}_0
           \mp \sqrt{ \Delta^2_0 - \Delta^{\text{c}\,2}_0 }\: \right )/ 2 \;.
  \end{equation}
For $n_h > 1.0$,
$\Delta_{\text{eff}}$ tends to zero in the limit
$\Delta_0/|t_{pd}| \to \infty$.
This is because
the renormalization factor $q$ tends to zero
while the ratio $\alpha$ defined in Eq.\ (\ref{eq:alpha})
has to be smaller than a finite value which is
determined by Eq.\ (\ref{eq:ndga}) with the condition $n_d = 1.0$.
On the other hand, in the electron-doped cases
there is no upper bound in $\alpha$,
and $q>0.0$.
Thus, $\Delta_{\text{eff}}$ increases with $\Delta_0$ for \,$n_h < 1.0$.
The charge excitation gap $\epsilon_{\text{gap}}$, defined by a discontinuity
of the chemical potential, can be also obtained analytically as
  \begin{eqnarray}
  \epsilon_{\text{gap}} &=&
       \left. {dE_{\text{GA}}^{(\text{o})} \over dN_h } \right |_{n_h=1 + 0}
   - \; \left. {dE_{\text{GA}}^{(\text{o})} \over dN_h } \right |_{n_h=1 - 0}
         \;, \nonumber \\
      & &  \\
      &=& \left \{
            \begin{array}{ll}
            0    & \quad  \mbox{for} \quad \Delta_0 < \Delta^{\text{c}}_0  \\
            2 \sqrt{ \Delta^2_0 - \Delta^{\text{c}\, 2}_0 }
                 & \quad \mbox{for}  \quad \Delta_0 \geq \Delta^{\text{c}}_0
              \end{array}
             \right. \;, \nonumber
  \end{eqnarray}
where $E_{\text{GA}}^{(\text{o})}$ is the minimized value of $E_{\text{GA}}$.
For $\Delta_0 \geq \Delta^{\text{c}}_0$, \,
$\epsilon_{\text{gap}}$ is finite and
equal to twice of the jump in $\Delta_{\text{eff}}$, i.e.,
$\epsilon_{\text{gap}} =
2 \,(\, \Delta_{\text{eff}}^{-0} - \Delta_{\text{eff}}^{+0}\,)$.

\subsection{Remarks}

The basic equations presented in this appendix,
Eqs.\ (\ref{eq:EGA})--(\ref{eq:Q}),
seems to be equivalent to those obtained by Balseiro {\em et al \/}
\cite{Balseiro} applying a functional integral approach due to
Kotliar and Ruckenstein \cite{Kotliar} to the three-band Hubbard model.
They have, however, neglected
the $\nu$ dependence of $\rho(\nu)$,
so that their results in the limit $U_d \to \infty$
differ from ours quantitatively.
The energy functional
obtained by Grilli {\em et al \/}
within an auxiliary-boson mean-field theory \cite{Grilli}
is also similar to that in the GA.
The renormalization factor in their theory, however, is different
from Eq.\ (\ref{eq:Qinf}),
and is equal to the number of the auxiliary-bosons, i.e.,
$q \Rightarrow 1 - n_d $.
Thus, the critical value $\Delta_0^{\text{c}}$ obtained
in auxiliary-boson mean-field theory
is a factor $\sqrt{2}$ smaller than Eq.\ (\ref{eq:d0c}),
which comes from the denominator of Eq.\ (\ref{eq:Qinf})
in the limit $n_d \to 1$.
Similar results have also been obtained by Hirashima {\em et al \/}
with the exact treatment of the local constraint of
no doubly occupied Cu sites
within the leading order of the $1/N$-expansion,\cite{Hirashima}
in which, specifically, the critical value  $\Delta_0^{\text{c}}$
is the same as that of the auxiliary-boson mean-field theory.

\begin{figure}
\caption{ The expectation value of
the fraction of $d$ holes $n_d(\widetilde \Delta)$
as a function of variational parameter $\widetilde \Delta$
(in units of $|t_{pd}|$)
for several values of the hole concentration $n_h$.
The circles ($\bullet$) denote the VMC results
in the Gutzwiller wave function $|\Psi(\widetilde \Delta)\rangle$,
and the solid lines denote the $d$-hole fraction
in the Fermi sea $|\Phi (\widetilde \Delta)\rangle$.  }
\label{fig:ndren}
\end{figure}

\begin{figure}
\caption{  The expectation value of the $p$-$d$ hopping energy,
 {\em per unit cell},
$E_{pd}(\widetilde \Delta)$ (in units of  $|t_{pd}|$)
for several $n_h$.
The circles ($\bullet$) denotes the VMC results in the GF,
and the lines denote the corresponding expectation values in
the FS. }
\label{fig:Epdren}
\end{figure}

\begin{figure}
\caption{  The value of $\widetilde \Delta$
which corresponds to the minimum of $E_{pd}(\widetilde \Delta)$
(in units of  $|t_{pd}|$).}
\label{fig:epdmin}
\end{figure}

\begin{figure}
\caption{ The expectation value of the total energy, {\em per unit cell},
$E(\widetilde \Delta)$
in the  half-filled case for $\Delta_0 = 0.5$, $1.0$, and $1.5$
(in units of  $|t_{pd}|$) with $\Delta_0 = (\varepsilon_p - \varepsilon_d )/2$.
The origin of energy is taken to be
$(\varepsilon_p + \varepsilon_d )/2 = 0$. }
\label{fig:egren}
\end{figure}

\begin{figure}
\caption{  The optimal value of the variational parameter
$\widetilde \Delta_{\text{opt}}$ (in units of  $|t_{pd}|$)
as a function of (a) $\Delta_0$, and (b) $n_h$.
The lines in (b) are just guide to the eye. }
\label{fig:dopt}
\end{figure}

\begin{figure}
\caption{ Variational results for the total energy, {\em per unit cell},
as a function of $\Delta_0$ for several $n_h$
(in units of \, $|t_{pd}|$).
The circles ($\bullet$) denote the VMC results,
and the solid lines denote the GA results. }
\label{fig:eg}
\end{figure}

\begin{figure}
\caption{ Variational results for
the fraction of $d$-holes $n_d$
as a function of $\Delta_0$ (in units of  $|t_{pd}|$).
The circles ($\bullet$) denote the VMC results,
and the solid lines denote the GA results. }
\label{fig:nd}
\end{figure}

\begin{figure}
\caption{Doping dependence of the VMC results:
(a) the total energy {\em per unit cell \/}
and (b)  the $d$-hole fraction, for several values of $\Delta_0$.
The lines are just guide to the eye.}
\label{fig:egnd}
\end{figure}

\begin{figure}
\caption{ The momentum distribution function
for holes in the bonding-band  $n(k)$
defined in Eq.\ (\protect\ref{eq:nk}):
\ (a) the half-filled case  ($n_h=144/144$)
with $\widetilde \Delta  = 0.1$, $1.0$, $10.0$,
and $25.0$ (in units of  $|t_{pd}|$)
which correspond to the optimal value for
$\Delta_0 = 0.3$, $1.0$, $2.8$, and $4.0$, respectively,
\ (b) the hole-doped case  ($n_h=156/144$)
with $\widetilde \Delta  = 0.1$, $1.0$, and $10.0$
which correspond to the optimal value for
$\Delta_0 = 0.4$, $1.1$, and $4.0$, respectively,
\ (c) the electron-doped case ($n_h=132/144$)
with $\widetilde \Delta  = 0.1$, $1.0$, and $10.0$
which correspond to the optimal value for
$\Delta_0 = 0.2$, $0.9$, and $3.7$, respectively,
\ (d) $k$-points for a $12 \times 12$ system
with the periodic-antiperiodic boundary condition,
where the solid and open circles denote, respectively, the occupied and
unoccupied states in $|\Phi (\widetilde \Delta)\rangle$
for $n_h = 1.0$. }
\label{fig:nk}
\end{figure}

\begin{figure}
\caption{  The GA results for
(a) $q$, and (b) $\Delta_{\text{eff}}$ as functions of $\Delta_0$
(in units of  $|t_{pd}|$) for several $n_h$.}
\label{fig:Q}
\end{figure}

\end{document}